\documentclass{article}
\usepackage[
  a4paper,
  margin=1in,
  headsep=4pt, 
]{geometry}
\usepackage{lastpage}
\usepackage{amsmath}
\usepackage{amssymb}
\usepackage{dsfont}
\usepackage{mathtools}
\usepackage{amsthm}
\usepackage[round]{natbib}
\newtheorem{theorem}{Theorem}
\renewenvironment{proof}[1][\proofname]{{\bfseries #1.}\\}{\qed}

\allowdisplaybreaks

\title{A Proper Scoring Rule for Validation of Competing Risks Models}
\author{Zoe Guan$^1$}
\date{%
    $^1$ Department of Epidemiology and Biostatistics, Memorial Sloan Kettering Cancer Center, New York, NY 10017\\%
}

\begin{document}

\maketitle

\begin{abstract}
    Scoring rules are used to evaluate the quality of predictions that take the form of probability distributions. A scoring rule is strictly proper if its expected value is uniquely minimized by the true probability distribution. One of the most well-known and widely used strictly proper scoring rules is the logarithmic scoring rule. We propose a version of the logarithmic scoring rule for competing risks data and show that it remains strictly proper under non-informative censoring. 
\end{abstract}

\section{Introduction}

A probabilistic forecast is a prediction that specifies a probability distribution over the set of possible outcomes. The quality of probabilistic forecasts is typically assessed using scoring rules (for an overview, see \cite{gneiting2007strictly}, Chapter 10 of \cite{parmigiani2009decision}, and \cite{dawid2014theory}). Given a set of outcomes $\mathcal{X}$ and a family of probability measures $\mathcal{P}$ over $\mathcal{X}$, a scoring rule is a loss function $s: \mathcal{X} \times \mathcal{P} \to \mathds{R} \cup [-\infty, \infty]$ that assigns a number $s(x, Q)$ to each combination of $x \in \mathcal{X}$ and $Q \in \mathcal{P}$.

A rational forecaster who believes that the true distribution is $P \in \mathcal{P}$ will report a forecast $Q \in \mathcal{P}$ that minimizes the expected score under $P$,
\begin{flalign}
S(P, Q) \coloneqq  E_P[s(X, Q)].
\end{flalign}
A scoring rule is proper if $S(P, P) \leq S(P, Q)$ for all $P, Q \in \mathcal{P}$, and it is strictly proper if $S(P, P) < S(P, Q)$ for $Q \neq P$. Strictly proper scoring rules are desirable because they encourage honesty (i.e. they encourage forecasters to report their true beliefs) and reward accuracy \citep{winkler1994evaluating}.  

One of the most well-known and widely used strictly proper scoring rules is the logarithmic scoring rule proposed by \cite{good1952rational}:
\begin{flalign}
s(x, Q) =  -\log(q(x))
\end{flalign}
where $q$ is the probability density or mass function corresponding to $Q$. There are many theoretical and empirical arguments supporting the use of the logarithmic scoring rule in various prediction problems \citep{winkler1969scoring, phillips1966conservatism, benedetti2010scoring}. Besides being strictly proper, the logarithmic scoring rule is also local, which means that it depends only on the predicted probabilities for observed events and does not use the predicted probabilities for unobserved events. Moreover, the logarithmic scoring rule discourages the forecaster from assigning extreme probabilities to very rare or very frequent events \cite{benedetti2010scoring}, which might be desirable in settings where overconfident predictions have serious consequences.

In survival or failure time analysis, interest lies in predicting the time until the occurrence of a specific event, which might not be fully observed due to censoring. \cite{dawid2014theory} described a proper scoring rule, called the survival score, for the classical survival analysis setting with a single event type and non-informative right censoring. The survival score gives rise to a variant of the logarithmic scoring rule as a special case. In this paper, we consider a competing risks setting where there are multiple mutually exclusive event types. We propose a logarithmic scoring rule for this setting and show that it remains strictly proper under non-informative right censoring.

\section{Competing Risks Notation}

Suppose there are $M$ competing causes of failure. Let $T$ be the time to failure and let $J \in \{1, \dots, M\}$ denote the cause of failure. $T$ is potentially subject to right censoring. Let $C$ be the censoring time, which is assumed to be independent of $(T, J)$. We observe $Y=\min(T, C)$ and $\Delta_j=I[T \leq C, J=j]$, the indicator for whether failure type $j$ is observed, for $j=1,\dots,M$. We assume the times are discrete, but similar results apply for continuous-time data, with minor notation changes.

Let $Q$ be a probability distribution for $(T, J)$ and $G$ a probability distribution for $C$. Let $f_{j,Q}(t) = Q(T = t, J=j)$, $F_{j,Q}(t) = Q(T \leq t, J=j)$, and $F_Q(t) = \sum\limits_{j=1}^M F_{j,Q}(t)$. These functions can depend on covariates, but for simplicity we omit them from the notation. The joint probability mass function for $(Y, \Delta_1, \dots, \Delta_M)$ is
\begin{flalign}
\pi_{Q, G}(Y=y, \Delta_1=\delta_1, \dots, \Delta_M=\delta_M) = \prod_{j=1}^M f_{j,Q}(y)^{\delta_j}  (1-F_Q(y))^{1-\delta}  G(C \geq y)^{\delta} G(C = y)^{1-\delta} 
\end{flalign}
where $\delta=\sum\limits_{j=1}^m \delta_j$.

\section{Scoring Rule and Proof of Propriety}

We define a logarithmic scoring rule that evaluates a probability distribution for $(T, J)$ against the observed data $(y, \delta_1, \dots, \delta_M)$. We show that this scoring rule is strictly proper.

\begin{theorem}
Given a probability distribution $Q$ for $(T, J)$, define
\begin{flalign}\label{lsr}
s((y, \delta_1, \dots, \delta_M), Q) \coloneqq -\sum_{j=1}^M \delta_j \log(f_{j,Q}(y)) - (1-\delta)\log(1-F_Q(y)).
\end{flalign}
This is a strictly proper scoring rule for the distribution of $(T, J)$. \\
\end{theorem}

When $M=1$, (\ref{lsr}) is equivalent to a special case of the survival score from Section 3.5 of \cite{dawid2014theory} that is obtained by setting $\psi(\lambda) = \lambda \log{\lambda}$. \\

\begin{proof}[Proof of Theorem 1]

Let $P$ and $Q$ be probability distributions for $(T, J)$. Let $G$ be a probability distribution for $C$. Define
\begin{flalign}
S_G(P, Q) \coloneqq E_{P,G}[s((Y, \Delta_1, \dots, \Delta_M), Q)].
\end{flalign}

We will show that for any choice of $G$, $S_G(P, Q)$ is uniquely minimized by $Q=P$. 

\begin{flalign*}
& S_G(P, Q) - S_G(P, P) \\
&= \sum_{y, \delta_1, \dots, \delta_M}  \pi_{P, G}(y, \delta_1, \dots, \delta_M) \left(s((y, \delta_1, \dots, \delta_M), Q) - s((y, \delta_1, \dots, \delta_M), P)\right) \\
&= \sum_{y, \delta_1, \dots, \delta_M}  \pi_{P, G}(y, \delta_1, \dots, \delta_M) \log \left( \frac{\prod_{j=1}^M f_{j,P}(y)^{\delta_j}  (1-F_P(y))^{1-\delta}}{\prod_{j=1}^M f_{j,Q}(y)^{\delta_j}  (1-F_Q(y))^{1-\delta}} \right) \\
&= \sum_{y, \delta_1, \dots, \delta_M}  \pi_{P, G}(y, \delta_1, \dots, \delta_M) \log \left( \frac{\prod_{j=1}^M f_{j,P}(y)^{\delta_j}  (1-F_P(y))^{1-\delta} G(C \geq y)^{\delta} G(C = y)^{1-\delta} } {\prod_{j=1}^M f_{j,Q}(y)^{\delta_j}  (1-F_Q(y))^{1-\delta} G(C \geq y)^{\delta} G(C = y)^{1-\delta} } \right) \\
&= \sum_{y, \delta_1, \dots, \delta_M}  \pi_{P, G}(y, \delta_1, \dots, \delta_M) \log \left( \frac{\pi_{P, G}(y, \delta_1, \dots, \delta_M) } {\pi_{Q, G}(y, \delta_1, \dots, \delta_M) } \right) \\
&= D_{KL}( \pi_{P,G} || \pi_{Q,G}) \\
&\phantom{==} \text{\small where $D_{KL}(p || q)$ denotes the Kullback-Leibler divergence from $p$ to $q$} 
\end{flalign*}

Kullback-Leibler divergence is non-negative and $D_{KL}(p || q)=0$ if and only if $p(x)=q(x)$ for all $x$ \citep{mackay2003information}, so $S_G(P, Q)$ is uniquely minimized by $Q=P$.

\end{proof}

\section*{Acknowledgements}
I would like to thank Giovanni Parmigiani for pointing me to relevant literature and providing helpful suggestions.

\bibliography{sample.bib}

\end{document}